\documentclass[twoside]{mhd}
\usepackage{graphicx}
\usepackage{bm}
\mhdhead{40}{1}{1}

\title{\uppercase{The Tayler instability at low\\ magnetic Prandtl numbers:\\
Chiral symmetry breaking and synchronizable helicity oscillations}}

\author{F.~Stefani\inst{*}, V.~Galindo, A.~Giesecke, N.~Weber, T.~Weier}

\institute{Helmholtz-Zentrum Dresden -- Rossendorf, Bautzner Landstr. 400, 
01328 Dresden, Germany} 

\begin{document}
\maketitle

\noindent
\textbf{Abstract:}

The current-driven, kink-type Tayler 
instability (TI) is a key ingredient of the 
Tayler-Spruit dynamo model for the generation of stellar 
magnetic fields, but is also discussed as a 
mechanism that might hamper the up-scaling of 
liquid metal batteries. Under some circumstances, the TI 
involves a helical flow pattern which goes along with some
$\alpha$ effect. Here we focus on the chiral symmetry 
breaking and the related impact on the $\alpha$ effect 
that would be needed to close the 
dynamo loop in the Tayler-Spruit model. For low magnetic 
Prandtl numbers, we observe intrinsic oscillations of 
the $\alpha$ effect. These oscillations serve then as the basis 
for a synchronized Tayler-Spruit dynamo model, which could 
possibly link the periodic tidal forces of planets with the 
oscillation periods of stellar dynamos.


\section{Introduction}
\label{sec:intro}

Current-driven instabilities have been known for a 
long time in plasma physics \cite{Goedbloed}. A case in 
point is the so-called $z$-pinch \cite{Bergerson_2006}, 
i.e. a straight 
current $j_z$ guided through the plasma which
produces an azimuthal magnetic field $B_{\varphi}$.
This field  is susceptible to both the axisymmetric ($m=0$)
sausage instability and the non-axisymmetric ($m=1$) kink 
instability.

In plasmas, the kink instability saturates by two processes 
which can be interpreted in terms of
mean-field MHD. First, the $\beta$ effect leads to 
a (radially dependent) 
counter-current which modifies the radial dependence 
$B_{\varphi}(r)$ so that the (ideal) stability condition 
$\partial (r B^2_{\varphi}(r))/\partial r<0$ \cite{Tayler_1973}
becomes marginally fulfilled. Second, the $\alpha$ effect   
leads to some azimuthal current $j_{\varphi}$, 
producing a $B_z$ component, which contributes to 
saturation according to the Kruskal-Shafranov
condition for the safety parameter.
However, the occurrence of a finite value of $\alpha$ is
not obvious, since it requires a spontaneous symmetry 
breaking between a left handed and a right handed TI 
mode which are, in principle, equally likely 
\cite{Gellert_2011,Bonanno_2012}.

While the magnetic Prandtl number 
${\rm Pm}=\mu_0 \sigma \nu $ of fusion plasmas is 
typically close to one, there are other 
relevant problems that are characterized by much
smaller  values. This applies, in particular, to 
liquid metal batteries \cite{Stefani_2011,Weber_2014,Stefani_2016a} 
whose upscalability 
might be limited 
by the kink-type Tayler instability (TI) 
\cite{Tayler_1973}. Using a quasi-stationary code 
\cite{Weber_2012} on the basis of the 
OpenFOAM library we have recently shown \cite{Weber_2015} 
that the saturation mechanism of the TI changes
completely for low $\rm Pm$: here, the quadratic combination 
of the $m=1$ velocity perturbations produces
$m=0$ and $m=2$ velocity components
which suppress the further growth of the TI.

Another interesting effect that was observed in 
\cite{Weber_2015} is the occurrence of helicity 
oscillations in the saturated state.
While those oscillations are not very 
interesting for liquid metal batteries,
they could be highly relevant for stellar dynamo models
of the Tayler-Spruit type \cite{Spruit_2002}.
The poloidal-to-toroidal field transformation 
for this type of nonlinear dynamo 
is easily provided by the usual
$\Omega$ effect due to 
differential rotation, but the
toroidal-to-poloidal field transformation
requires some $\alpha$ effect to be produced by the TI.
If this $\alpha$ effect has a tendency  to 
intrinsic oscillations, 
this could give a chance for weak external forces 
(such as exerted by planets) to synchronize the 
entire dynamo. 

This paper summarizes the corresponding ideas 
as outlined in \cite{Weber_2015} and 
\cite{Stefani_2016}, and adds a new aspect related 
to the question of whether this type of synchronized dynamo
might provide the correct orientation of the 
butterfly diagram for sunspots.

\section{Helicity waves in the saturated state of the Tayler instability}

For the sake of concreteness, we study the features of the 
TI in a cylinder of height-to-diameter ratio 
$H/(2R)=1.25$, which is passed through by an axial current. 
We choose ${\rm Pm}=10^{-6}$ and  
a Hartmann number ${\rm Ha}=B_{\varphi} R (\sigma/\rho \nu)^{1/2}=100$
which is already significantly higher than the critical
Hartmann number ${\rm Ha}_{\rm{crit}}=21.09$ for the infinitely long cylinder
\cite{Herreman_2015}.

\begin{figure}
  \centering
  \includegraphics[width=0.99\textwidth]{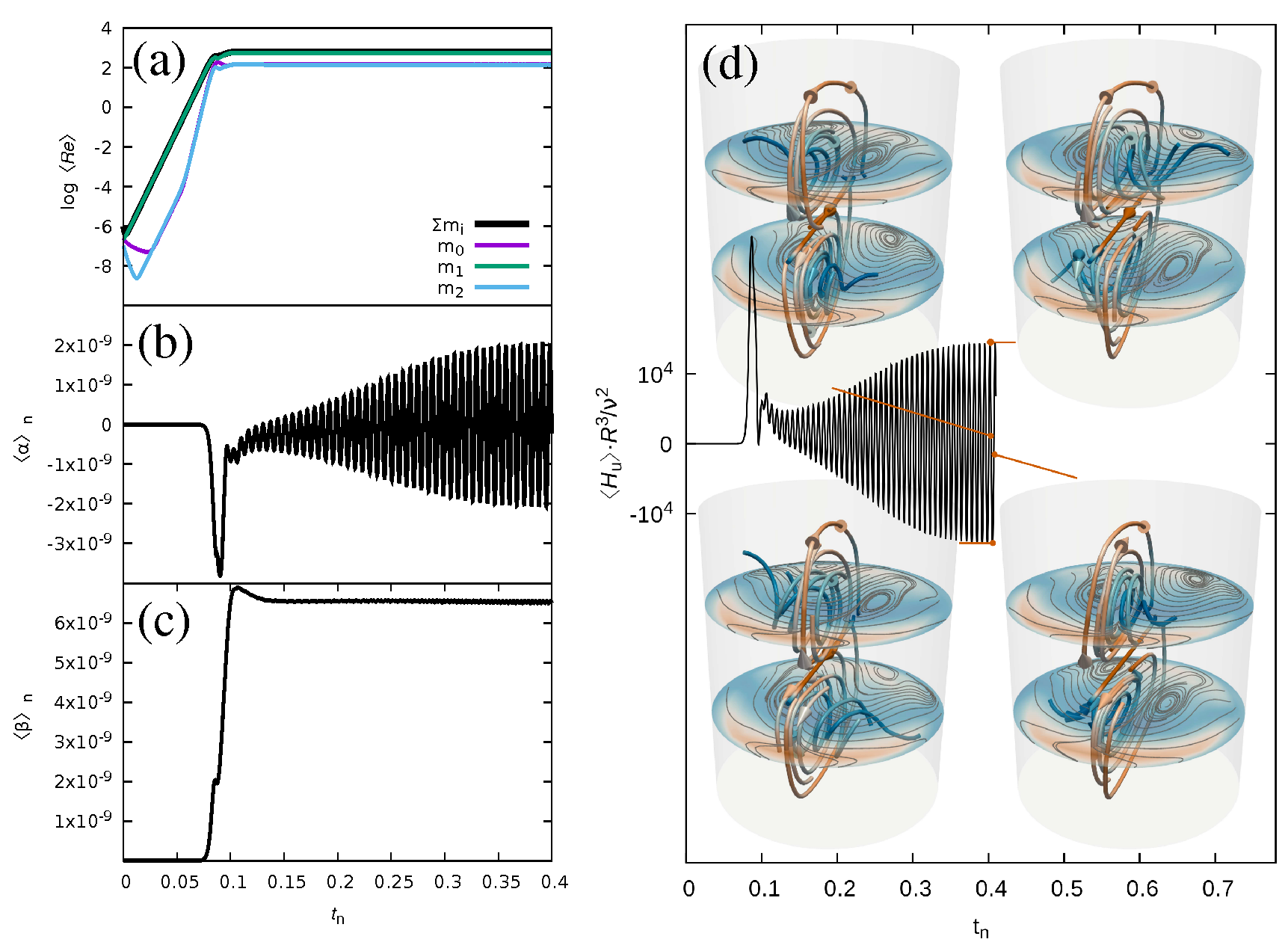}
  \caption{Evolution of various quantities for a developing TI at
   ${\rm Pm}=10^{-6}$ and ${\rm Ha}=100$. (a) Reynolds number of the
   total flow, and of the first azimuthal modes individually.
   The exponential increase (note the log-scale of the 
   ordinate axis) is followed by a saturated state with a nearly constant
   Reynolds number. (b) Normalized $\alpha$ effect, showing 
   a spontaneous symmetry breaking in the kinematic growth 
   phase of the TI, and a clear oscillation in the saturated phase.
   (c) Normalized $\beta$ effect, showing a rather constant value
   in the saturated regime. 
   (d) Normalized helicity, and four snapshots of the velocity field
   in the saturation regime. Note the slightly changing tilts of the 
   two vortices which produce the oscillation of the 
   helicity and $\alpha$.
   After \cite{Weber_2015}.
  }
  \label{fig:1}
\end{figure}

Figure~1 illustrates the occurrence of helicity oscillations 
in the saturated regime of the TI. Panel (a)  
shows the logarithms of the Reynolds number for the 
total flow 
and its individual azimuthal modes $m=0,1,2$.
The initial exponential increase of the dominant $m=1$ mode
is later accompanied by the steeper increase of the 
nonlinearily produced $m=0$ and $m=2$ modes. Saturation sets
in shortly before $t_n=0.1$ (the time is normalized to 
the viscous time scale) 
when the energy of the
$m=0$ and $m=2$ modes becomes comparable to that of the dominant
$m=1$ mode. What is interesting is the behavior of the
$\alpha$ effect in panel (b), which produces an azimuthal current 
and, therefore, an axial  magnetic field. 
In the saturated regime (for $t_n>0.1$), this 
spontaneous symmetry breaking gives way 
to a pronounced oscillatory behaviour. 
In contrast to the oscillation of $\alpha$ and the related helicity, 
the $\beta$ effect (c)  is rather constant in the saturation regime.

An illustration of the helicity oscillation is provided in Fig.~1d
which shows the velocity field of the TI at 4 particular instants.
While the two main TI vortices, which are typical for the chosen
aspect  ratio, point essentially in the same direction, we can observe 
some slight change of their tilt which produces 
the oscillation of the helicity and $\alpha$.

\section{Synchronized helicity oscillations and the solar dynamo cycle}

The observed helicity oscillation seems to be 
a generic feature of the saturated TI for large enough 
$\rm Ha$ at low $\rm Pm$ (note that Bonanno and Guarnieri 
\cite{Bonanno_2016} could not 
find helicity oscillations in a dissipation-free model).
We ask now for its possible implications for Tayler-Spruit type
stellar dynamos. Specifically, we will introduce some 
$m=2$ viscosity perturbation of the base
state which serves as a surrogate for the tidal torque of planets on the 
stellar tachocline. The background of our consideration is 
the claimed relation  
of the $\sim 11$ years period of the dominant tidal forces of the 
Venus-Earth-Jupiter system with the $\sim 22$ years of the solar cycle
\cite{Wolf_1859,Bollinger_1952,Takahashi_1968,Wood_1972,Hung_2007,Wilson_2013,Okhlopkov_2014}.

Although these tidal forces are usually considered as much too weak
to influence the solar dynamo, one should also keep in mind 
the large gravitational acceleration at the tachocline 
that amounts to 540 m/s$^2$ \cite{Wood_2010}. This translates 
the apparently tiny  tidal heights in the order 
of 1 mm \cite{Condon_1975} 
to  equivalent velocities of $v=(2 g h)^{1/2}\sim 1 $\,m/s.
Such velocities, when allowed to 
coherently develop in the quiet regions of the tachocline (and not 
being overwhelmed by the highly fluctuating velocities 
prevailing in the convection zone) might indeed be relevant for the 
dynamo.

\begin{figure}
  \centering
  \includegraphics[width=0.8\textwidth]{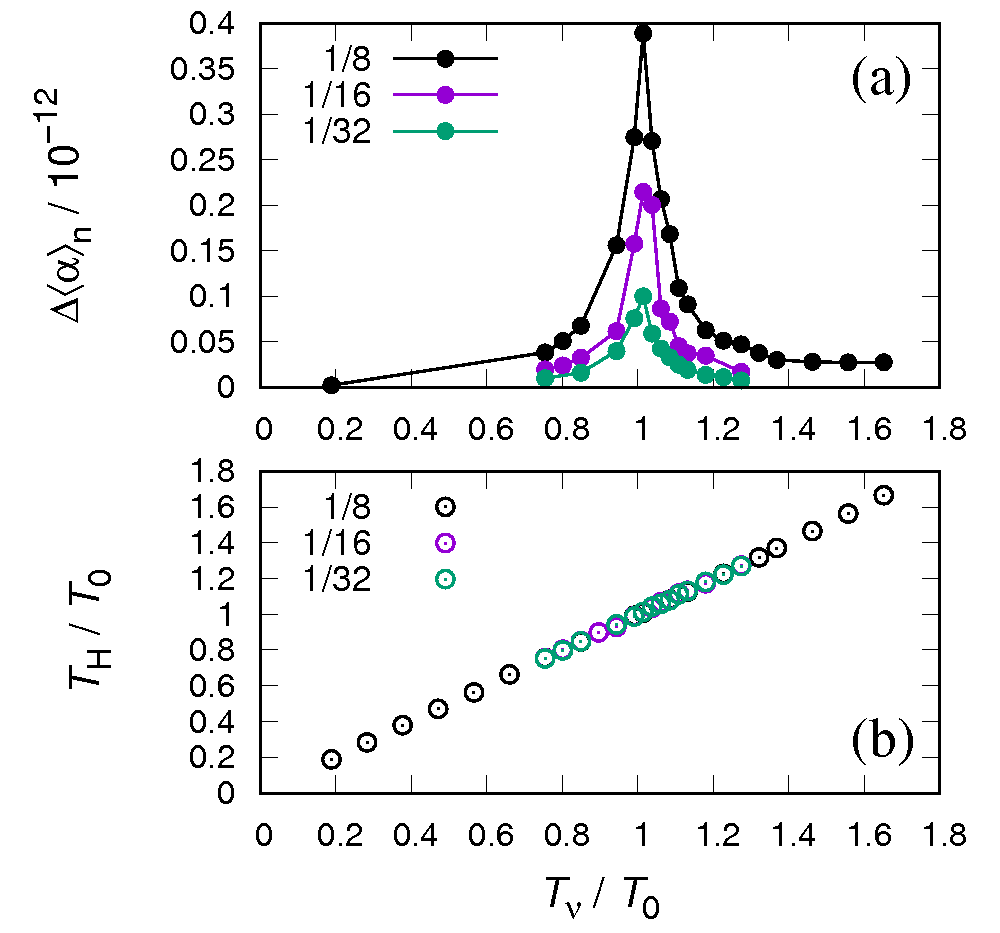}
  \caption{Resonance between an $m=2$ viscosity oscillation according 
  to Eq. (1) and the oscillation of $\alpha$.
  (a) Amplitude of the $\alpha$ oscillation in dependence on
   $T_{\nu}/T_0$. (b) Ratio $T_H/T_0$ in dependence 
   on $T_{\nu}/T_0$, showing  a 
   clear 1:1 resonance. After \cite{Stefani_2016}.
  }
  \label{fig:2}
\end{figure}

Without any perturbation, when choosing ${\rm Ha}=80$,
the TI would lead to a weak helicity oscillation with a 
certain period $T_0$. We impose on this state 
an $m=2$ oscillation  of the viscosity $\nu$
in the form 
\begin{eqnarray}
\nu(r,\phi,t)&=&\nu_0 \{  1    +  A [   1+ 0.5 r^2/R^2 \sin(2 \phi) (1+  \cos(2 
\pi t/T_{\nu}) )     ]   \} 
\end{eqnarray}
which includes a constant term $\nu_0 (1+A)$ and an additional 
term with an $m=2$ azimuthal dependence that is oscillating with
a period $T_\nu$. 
For the intensity of the viscosity wave we select now 
three specific values $A=1/32,1/16,1/8$.
The resulting amplitude of the oscillation of $\alpha$ (and the helicity) 
is shown
in Fig.~2a, its period $T_H$ in Fig.~2b.  Obviously, we obtain
a strong 1:1 resonance of the oscillation amplitude at 
$T_{\nu}=T_0$.

What are the possible consequences of this synchronization of $\alpha$ 
for a complete Tayler-Spruit dynamo? To answer this question 
we consider the simple zero-dimensional equation system
\begin{eqnarray}   
    \frac{d a(t) }{d t} &=& \alpha(t) b(t) - \tau^{-1} a(t)\\
    \frac{d b(t)}{d t} &=& \Omega a(t) - \tau^{-1} b(t)
   \end{eqnarray}
which describes the transformation of poloidal field $a$ to 
toroidal field $b$ via some $\Omega$ effect, and the back-transformation
of toroidal to poloidal field via the TI-based $\alpha$ effect.
The free decay time of the respective modes is denoted by
$\tau$. Note that similarly simple equation systems 
have been widely used to understand various dynamo features 
\cite{Wilmot_2006}.
Hereby, $\alpha(t)$ is parametrized according to
\begin{eqnarray}   
    \alpha(t) &=& \frac{c}{1+g b^2(t)}
                          +\frac{p b^2(t)}{1+ h b^4(t)} \sin{(2 \pi t/T_{\nu})} 
	\label{alpha}		  
   \end{eqnarray}
which represents, in its first part, some constant term which
is only quenched by magnetic field energy $b^2$, plus a time-dependent
part with the tidal period $T_{\nu}$, the prefactor of which is 
chosen such as to emulate the resonance condition $T_\nu=T_0$
seen in Fig.~2.
\begin{figure}
  \centering
  \includegraphics[width=0.8\textwidth]{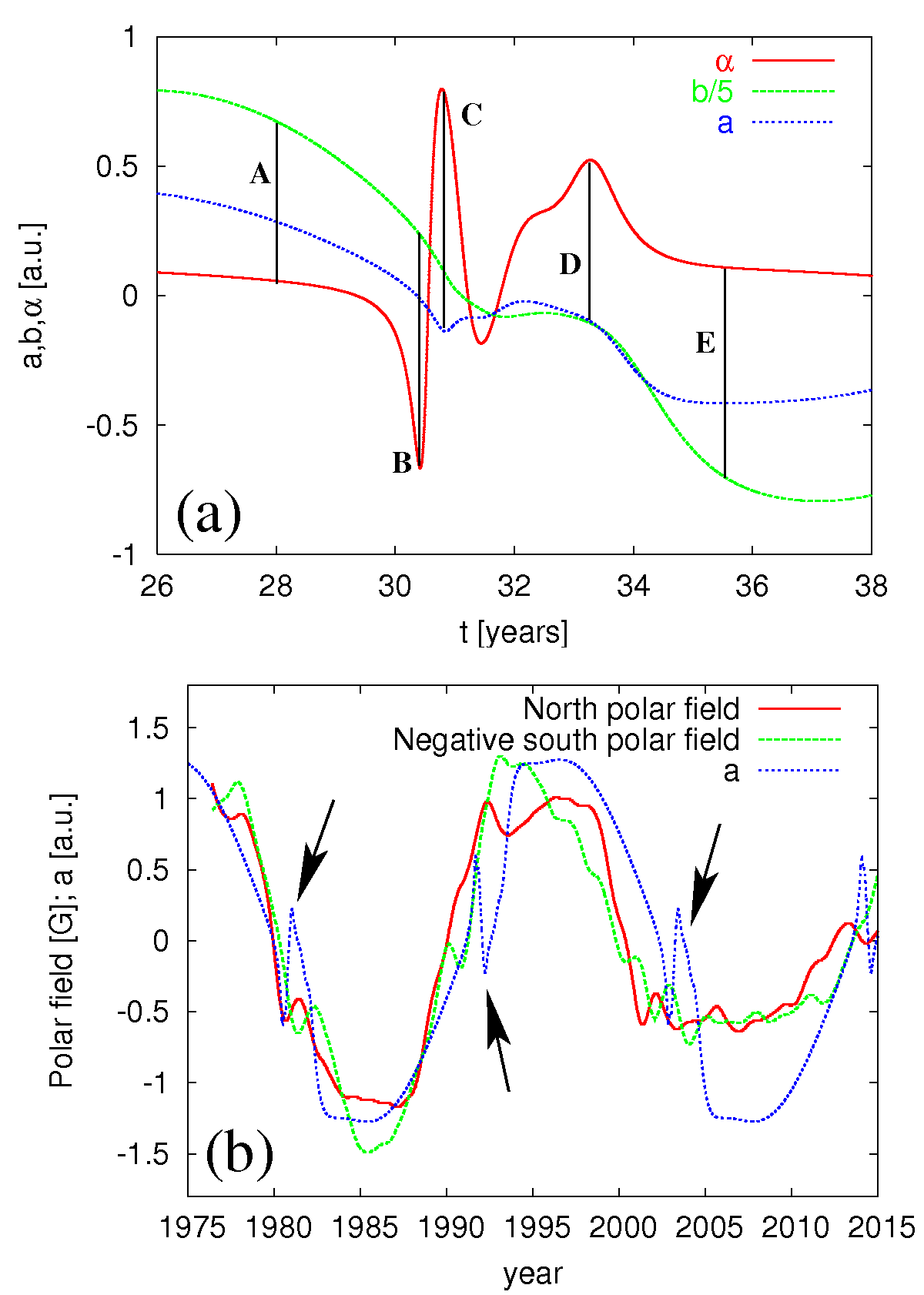}
  \caption{Evolution of the equation system (2-4) with 
	     tidal forcing of 11.07 years. 
	     (a) Parameter choice $c=0.8$, $g=1$, 
	      $p=8$, $h=10$, $\Omega=10$. 
	      (b) Comparison of the north and south polar magnetic 
	       field and the parameter $a$, computed 
	       for $c=0.8$, $p=8$, $h=10$, $\Omega=50$, 
	       $g=1$, appropriately scaled and shifted in time. 
	       The field data are the 20 nHz filtered data from Wilcox 
	       Solar Observatory (courtesy J.T. Hoeksema).
	       After \cite{Stefani_2016}.
  }
  \label{fig:3}
\end{figure}

The resulting dynamo behaviour is quite interesting. 
Figure 3a shows approximately half a period of a dynamo
cycle
for the particular choice $c=0.8$, $g=1$, $p=8$, $h=10$, $\Omega=10$.
We observe a clear sign change of the magnetic field, 
and amazing ''spiky'' 
features of $\alpha$ close to the turning point of $a$ and $b$. 
The capitals A...E mark various instants with specific 
features to be explained in the following:
Initially, at A, $\alpha$ is
strongly quenched by 
the large value of $b$, while its
oscillatory part is negligible since $b$ is so strong that 
we are far away from resonance. While $b$ decreases,
it reaches a level at which  
the TI helicity oscillation becomes resonant with the
viscosity oscillation. This happens at B 
when $b\sim 0.56$, which 
actually corresponds to 
the maximum of the pre-factor $b^2/(1+10 b^4)$ of the 
oscillatory term in Eq. (4).
At this point $\alpha$ becomes strongly negative.
Shortly after, at C, $b$ drops to zero,
so that the quenching of the constant term  of $\alpha$  
disappears and $\alpha$ acquires the unquenched value $c$ (here
$\sim 0.8$). Later, at D, $b$ passes again through
the resonant point $b\sim-0.56$ for the helicity oscillation 
so that the oscillatory part contributes again its large, 
but now positive, value to $\alpha$. Finally, at E, $b$ 
increases quite 
smoothly until it reaches a maximum amplitude where 
$\alpha$ is strongly quenched and rather constant.

In Fig. 3b we compare an appropriately scaled and
time-shifted segment of our $a(t)$ with the available 
time series of the 20 nHz filtered 
north and south polar field data. 
An amazing coincidence exists between the additional 
peaks of the
north and south polar field in Fig. 3(b) and 
the corresponding spikes of our $a$ (indicated by the three black
arrows). 
Another point, seen in Fig. 3(a), 
is related to the vigorous, ''spiky'' 
variations  of $\alpha$ close to the
reversal point of $a$ and $b$. It is tempting 
to relate this  behaviour
to the short-term sign changes of the current-helicity, as 
observed recently by \cite{Zhang_2012}.

\section{And the butterfly diagram?}

For decades, mean-field theory had served as the standard
model of the solar dynamo, providing a natural 
explanation for the periodicity and the
equator-ward sunspot propagation of the solar cycle 
\cite{Steenbeck_1969}. However, this model suffered a blow when 
helioseismology mapped the differential rotation
in the solar interior \cite{Brown_1989}. 
In particular, the positive radial shear in a
±30$^{\circ}$ strip around the equator results in a 
serious problem with the Parker-Yoshimura sign
rule that requires $\alpha \partial \Omega/\partial r <0$   
in the 
northern hemisphere for the correct equator-ward
propagation of sunspots \cite{Parker_1955,Yoshimura_1975}. 

A possible solution of this dilemma was found in 
the Babcock-Leighton mechanism
\cite{Babcock_1961,Leighton_1964}, which interprets 
the generation of poloidal field by the
stronger diffusive cancellation of the leading 
sunspots (closer to the equator) compared
with that of the trailing spots (farther from the equator). 
This leads to a spatially separated,
or flux-transport, type of dynamo 
\cite{Choudhuri_1995}, which also
provides the correct butterfly diagram when it is combined 
with an appropriate meridional
circulation.

Coming back to our synchronization model, we might
ask what direction of the butterfly diagram it would provide.
For this purpose we consider the following slight extension of the
equation system (2-4):
\begin{eqnarray}   
    \frac{d a_1(t) }{d t} &=& \alpha(t) b_1(t) - \tau_{a_1}^{-1} a_1(t)\\
    \frac{d a_2(t) }{d t} &=& \alpha(t) (b_1(t)+b_2(t)) - \tau_{a_2}^{-1} a(t)\\
    \frac{d b_1(t)}{d t} &=& \Omega (a_1(t)- \kappa a_2(t)) - \tau_{b_1}^{-1} b_1(t)\\
     \frac{d b_2(t)}{d t} &=& \kappa \Omega a_2(t) - \tau_{b_2}^{-1} b_2(t) \; .   
       \end{eqnarray}
Here, the amplitudes $a_1$ and $a_2$ represent the first and the second
possible meridional harmonics of the
poloidal field, while $b_1$ and $b_2$ stand for the
first two harmonics of the toroidal field.
Although this equation system is motivated by 
the system given by Nefedov and Sokoloff \cite{Nefedov_2010} 
we keep its pre-factors a bit more generic since we 
actually do not know the exact meridional dependence 
of the first two relevant (poloidal and toroidal) eigenmodes.

While in the original paper \cite{Nefedov_2010} the free 
decay rates of the individual modes were derived as
$\tau^{-1}_{a_1}=1$, $\tau^{-1}_{a_2}=9$, $\tau^{-1}_{b_1}=4$, and
$\tau^{-1}_{b_2}=16$ (and the factor $\kappa=3$),
we choose here $\tau^{-1}_{a_1}=1$, 
$\tau^{-1}_{a_2}=3.2$, $\tau^{-1}_{b_1}=1$, and
$\tau^{-1}_{b_2}=3.2$, $\kappa=1$.
 
The resulting butterfly diagram is shown in Fig. 4, when further
choosing $c=0.8$, $g=1$, $p=8$, $h=10$, $\Omega=100$, and 
$T_0=11.07$ yr. Surprisingly, our synchronization model  
produces the correct orientation of the butterfly diagram even 
when the product of $c$ (the constant part of $\alpha$ )
with $\Omega$ is positive! However, given the many parameters that
enter the equation system (5-8), together with the 
complex expression for $\alpha$, more parameter studies 
are necessary in order to 
check the robustness  of this behaviour.

\begin{figure}
  \centering
  \includegraphics[width=0.9\textwidth]{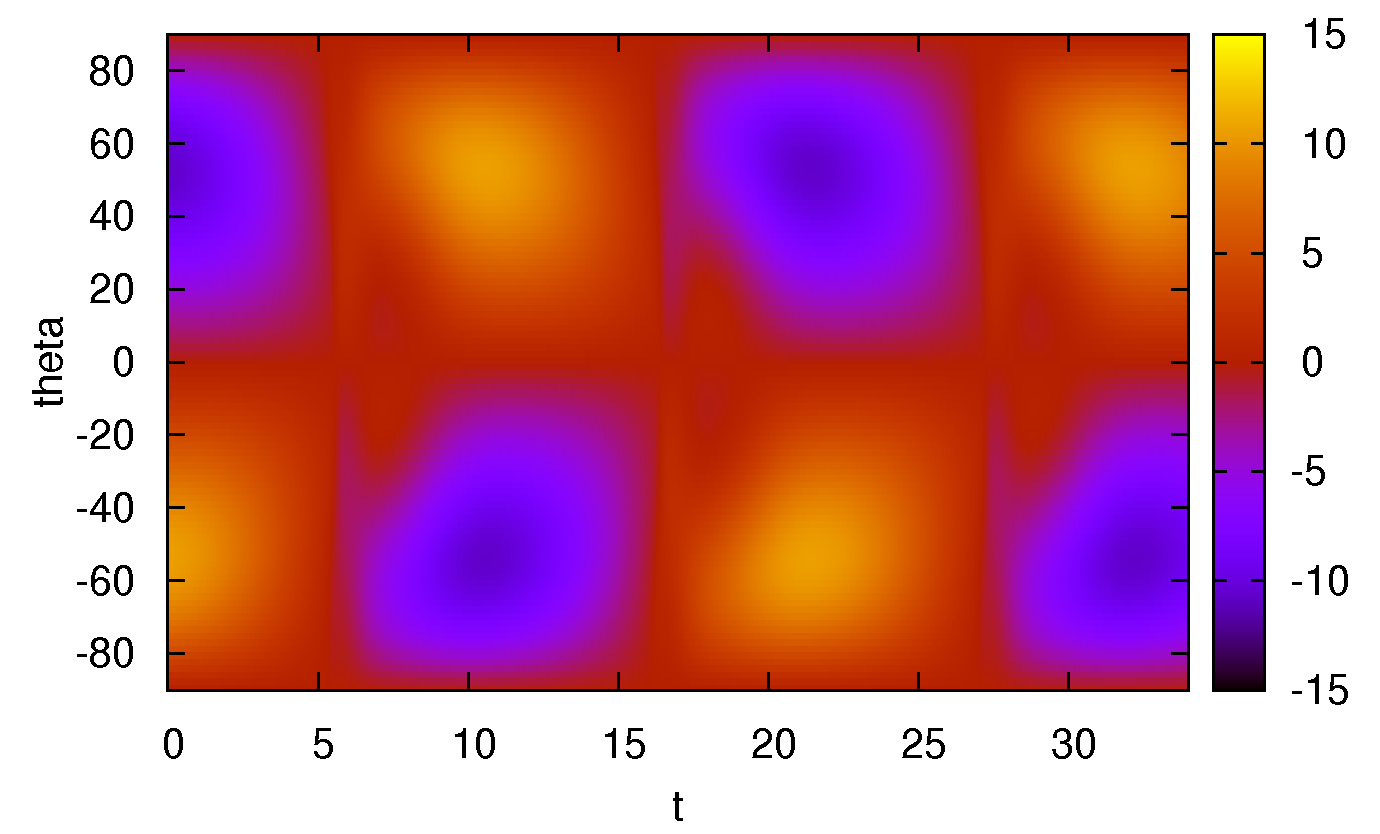}
  \caption{
  Butterfly diagram resulting from the equation system (5-8) when choosing
  $\tau^{-1}_{a_1}=1$, $\tau^{-1}_{a_2}=3.2$, $\tau^{-1}_{b_1}=1$, and
$\tau^{-1}_{b_2}=3.2$, $\kappa=1$, and
$c=0.8$, $g=1$, $p=8$, $h=10$, $\Omega=100$, and 
$T_0=11.07$ yr. Note that, although $c$, i.e. the constant part of $\alpha$, 
 and $\Omega$ are both positive, the butterfly is oriented in the right 
  direction.}
  \label{fig:4}
\end{figure}


\section{Conclusions}

While the traditional explanation of the Hale cycle 
of the solar magnetic field relies on intrinsic 
features of the solar dynamo such
as the magnetic diffusivity, the 
amplitudes of $\Omega$, $\alpha$ and the meridional flow 
\cite{CharbonneauDikpati_2000},
we have focused on a  mechanism that could allow
for synchronizing the solar dynamo with planetary tides. 

Motivated by the spontaneous occurrence of 
helicity oscillations in the saturated state of 
the TI as observed in 
\cite{Weber_2015}, we studied a simplified model for the resonant 
excitation of those oscillations by a viscosity 
oscillation with $m=2$ azimuthal dependence that
serves as dummy for a tidal forcing.  
The helicity and $\alpha$ oscillations, thought to 
be excited by
the 11.07\,years periodic tide produced by the
Venus-Earth-Jupiter system,
served as a ``clock'' for the 22.14\,years dynamo cycle 
of a reduced, zero-dimensional 
$\alpha-\Omega$ dynamo model. 
Actually, similar resonance phenomena have been discussed 
in connection with the swing excitation of galactic 
dynamos \cite{Chiba_1990} and with the von-K\'arm\'an-sodium
dynamo experiment \cite{Giesecke_2012}.
However, it is
a key feature of the mechanism discussed here that
it requires only weak external perturbations to 
trigger the helicity oscillations which just 
reshuffle the energy between left and right handed TI modes. 
That way the tiny planetary forces might indeed get  
a chance to synchronize the solar dynamo.
By slightly extending the model of \cite{Stefani_2016}
we have also shown that our model may lead to the
correct orientation of the butterfly diagram.

An obvious next step is related to the 
question of whether longer periodicities of the solar 
dynamo, such as the 87-year
Gleissberg cycle, the 210-year Suess-de-Vries cycle, and 
the 2300-year Hallstatt cycle \cite{Abreu_2012,Charvatova_1997,Jose_1965,
Palus_2000,Scafetta_2014,Scafetta_2016},
could also be explained in the framework of the  
synchronization model.
The solution of this problem
might have tremendous consequences,
in particular, if any of the disputed mechanisms
for connecting solar activity and terrestrial 
climate 
\cite{Svensmark_1997,Scafetta_2010,Gray_2010,Gervais_2016} 
could be validated.

\section*{Acknowledgments}
This work was supported by the Deutsche Forschungsgemeinschaft 
in the frame of the SPP 1488 (PlanetMag), as well as by 
Helmholtz-Gemeinschaft Deutscher Forschungszentren (HGF)
in frame of the Helmholtz alliance LIMTECH.
Wilcox Solar Observatory data used in this study 
was obtained via the web site 
wso.stanford.edu 
(courtesy of J.T. Hoeksema).
F. Stefani thanks R. Arlt, A. Bonnano, A. Brandenburg, A. Choudhuri, 
D. Hughes, M. Gellert, G. R\"udiger, and 
D. Sokoloff for fruitful discussion on the solar-dynamo mechanism.
We thank V. Pipin for pointing out a wrong axis labeling 
in Fig.~7 of \cite{Stefani_2016} which is now corrected in
Fig.~3b.





\lastpageno	


\end{document}